\documentclass[10pt,conference,a4paper]{IEEEtran}
\IEEEoverridecommandlockouts								
														  
\def\BibTeX{{\rm B\kern-.05em{\sc i\kern-.025em b}\kern-.08em
		T\kern-.1667em\lower.7ex\hbox{E}\kern-.125emX}}

\usepackage[utf8]{inputenc}

\usepackage{amsmath,amssymb,amsfonts,amsthm}
\DeclareMathOperator*{\argmin}{arg\,min}

\usepackage{color}
\usepackage{float}
\usepackage{mathtools}
\usepackage{textcomp}
\usepackage{soulutf8} 
\usepackage{xcolor}

\usepackage{tabularx}
\usepackage{soul} 
\usepackage{breqn}
\usepackage{multirow}

\usepackage{graphics} 
\usepackage{epsfig} 
\usepackage{graphicx}

\usepackage{algorithmicx}
\usepackage{algorithm}
\usepackage{algpseudocode}

\usepackage{verbatim}
\usepackage{subfig}

\newcolumntype{P}[1]{>
{\centering\arraybackslash}p{#1}}

\algnewcommand\algorithmicinput{\textbf{Input:}}
\algnewcommand\Input{\item[\algorithmicinput]}

\algnewcommand\SET[2]{\item\algorithmicset\ #1 \algorithmicto\ #2}

\makeatletter
\def\BState{\State\hskip-\ALG@thistlm}
\makeatother

\def\BibTeX{{\rm B\kern-.05em{\sc i\kern-.025em b}\kern-.08em
    T\kern-.1667em\lower.7ex\hbox{E}\kern-.125emX}}

\theoremstyle{remark}

\theoremstyle{definition}

\definecolor{orange}{RGB}{255,127,0}
\definecolor{gold}{rgb}{0.85,.66,0}

\newcommand{\cplxgauss}[2]{\mathcal{N}_{\mathbb{C}}(#1\,#2)} 
\newcommand{\uniform}[2]{\mathcal{U}(#1\,#2)} 

\newcommand{\eye}[1]{\mathbf{I}_{#1}} 
\newcommand{\zeros}[1]{\mathbf{0}_{#1}} 
\newcommand{\ones}[1]{\mathbf{1}_{#1}} 
\newcommand{\herm}{^{\scriptscriptstyle\mathrm{H}}} 
\newcommand{\transp}{^{\scriptscriptstyle\mathrm{T}}} 
\newcommand{\abs}[1]{\left\lvert#1\right\rvert} 
\newcommand{\norm}[1]{\left\lVert#1\right\rVert} 
\newcommand{\trace}[1]{\mathrm{tr}\left(#1\right)} 
\newcommand{\diag}[1]{{\mathrm{diag}\left({#1}\right)}}

\newcommand{\inreal}[1]{{\ \in \mathbb{R}^{#1}}}
\newcommand{\incplx}[1]{{\ \in \mathbb{C}^{#1}}}

\newcommand{\sindex}{^{(s)}}

\newcommand{\nscovmatrix}{\boldsymbol{\Theta}_{k}}

\newcommand{\sinrks}{\gamma\sindex_{k}}
\newcommand{\vvectors}[1]{\mathbf{v}\sindex_{#1}}
\newcommand{\hvectors}[1]{\mathbf{h}\sindex_{#1}}

\newcommand{\BBmatrix}{\mathbf{B}\sindex}
\newcommand{\yHats}{\hat{\mathbf{y}}_{0}\sindex}

\newcommand{\meanK}{\bar{K}\sindex}

\title{Low-Complexity Distributed XL-MIMO \\ for Multiuser Detection}

\author{
	\IEEEauthorblockN{Victor Croisfelt Rodrigues\IEEEauthorrefmark{1}, Abolfazl Amiri\IEEEauthorrefmark{2},  Taufik Abr\~{a}o\IEEEauthorrefmark{3}, Elisabeth de Carvalho\IEEEauthorrefmark{2}, and Petar Popovski\IEEEauthorrefmark{2}}\\
	\IEEEauthorblockA{\IEEEauthorrefmark{1}Universidade de S\~{a}o Paulo, S\~{a}o Paulo, Brazil, \\ \IEEEauthorrefmark{2}Aalborg University, Aalborg, Denmark\\ \IEEEauthorrefmark{3}Universidade Estadual de Londrina, Londrina, Brazil\\
	E-mail: victorcroisfelt@usp.br, \{aba,edc,petarp\}@es.aau.dk and taufik@uel.br}
}

\begin{document}
\maketitle
\pagestyle{empty}

\begin{abstract}
In this paper, the zero-forcing and regularized zero-forcing schemes operating in crowded extra-large MIMO (XL-MIMO) scenarios with a fixed number of subarrays have been emulated using the randomized Kaczmarz algorithm (rKA). For that, non-stationary properties have been deployed through the concept of visibility regions when considering two different power normalization methods of non-stationary channels. {We address the randomness design of rKA based on the exploitation of spatial non-stationary properties.} Numerical results show that, in general, the proposed rKA-based combiner {applicable to} XL-MIMO systems can considerably decrease computational complexity of the {signal detector} by paying with small performance losses.
\end{abstract}

\section{Introduction}
Motivated by higher area throughput that extremely large arrays can offer \cite{Bjornson2019},
recent notable research efforts are being carried out to improve the scalability of the so-called extra-large MIMO (XL-MIMO) systems. Due mainly to increased spatial resolution and the emergence of non-stationary channels, this new vision is currently materializing as an important beyond 5G technology and being considered as a distinct operating regime of Massive MIMO (M-MIMO) \cite{DeCarvalho2019}. With physical large arrays, 
spatial non-stationarity and inherent high array dimensions under user crowded scenarios have significant {harmful} impacts on the performance and computational complexity of linear receive combining techniques{, which are traditionally used in M-MIMO systems \cite{Bjornson2017}}. {This calls for} different manners of performing receive combining in XL-MIMO systems, {which try to exploit non-stationarities and seek a good trade-off} between performance and computational complexity when a large number of users are served.

{Taking into account crowded scenarios and the desire for low cost base stations (BSs),} several low-complexity linear detection algorithms that attempt to relax the computation of known linear receive combining criteria have been proposed in recent years for canonical M-MIMO; such as \cite{mller2013linear,Boroujerdi2018} to cite a few. These works, however, do not consider non-stationary channels that appear when antenna arrays are scaled up{, as is the case of XL-MIMO}. Meanwhile, the authors in \cite{Amiri2019} proposes a variational message passing (VMP) based symbol detection method for XL-MIMO and under crowded scenarios. Although the proposed method outperforms linear receivers, the algorithm demands the optimization of a damping factor, which accelerates the convergence of the algorithm, but unfortunately translates into undesired additional complexity. In addition to that, its complexity depends on the modulation order used to transmit user messages, making the comparison with linear receivers cumbersome. To the best of our knowledge, few are the works that study low-complexity linear receive combining techniques under the presented scenario of interest. 

\noindent{\textbf{Contributions}}: Inspired by the {promising} results obtained for M-MIMO \cite{Boroujerdi2018,Boroujerdi2019,Rodrigues2019}, this work proposes the application of the randomized Kaczmarz algorithm (rKA) as a way to circumvent the high-dimensional matrix inversion that comes with zero-forcing (ZF) and regularized zero-forcing (RZF) schemes when these are applied to recover the signal estimates of a crowded XL-MIMO scenario. The contributions are {listed as follows}: (i) extension of rKA to resemble the performance of ZF and RZF schemes for a XL-MIMO system with a fixed number of subarrays; (ii) consideration of non-stationary properties through the concept of visibility regions (VRs) when considering two different power normalization methods of non-stationary channels \cite{Ali2019}; (iii) exploitation of non-stationary features in the randomness design of rKA; (iv) {complexity analysis considering the different random variants of the proposed algorithm.}

{Some valuable features of the algorithm are as follows. \textit{Simplicity:} the only tuning parameter needed to be set is the number of iterations at each subarray. The others stem from network design choices and environment characteristics, which obviously affect the convergence of the algorithm, as discussed in \cite{Boroujerdi2018}, \cite{Boroujerdi2019}, and \cite{Rodrigues2019}. However, this also means that a convergence analysis is sufficient to characterize the efficiency of the algorithm to achieve its goal. \textit{{Graceful degradation:}} given {the computational constraints for any BS, we can
flexibly trade off the number of iterations with the performance.} 
}

\section{System Model}\label{sec:sysmodel}
\begin{figure}[h]
	\centering
	\includegraphics[width=.47\textwidth]{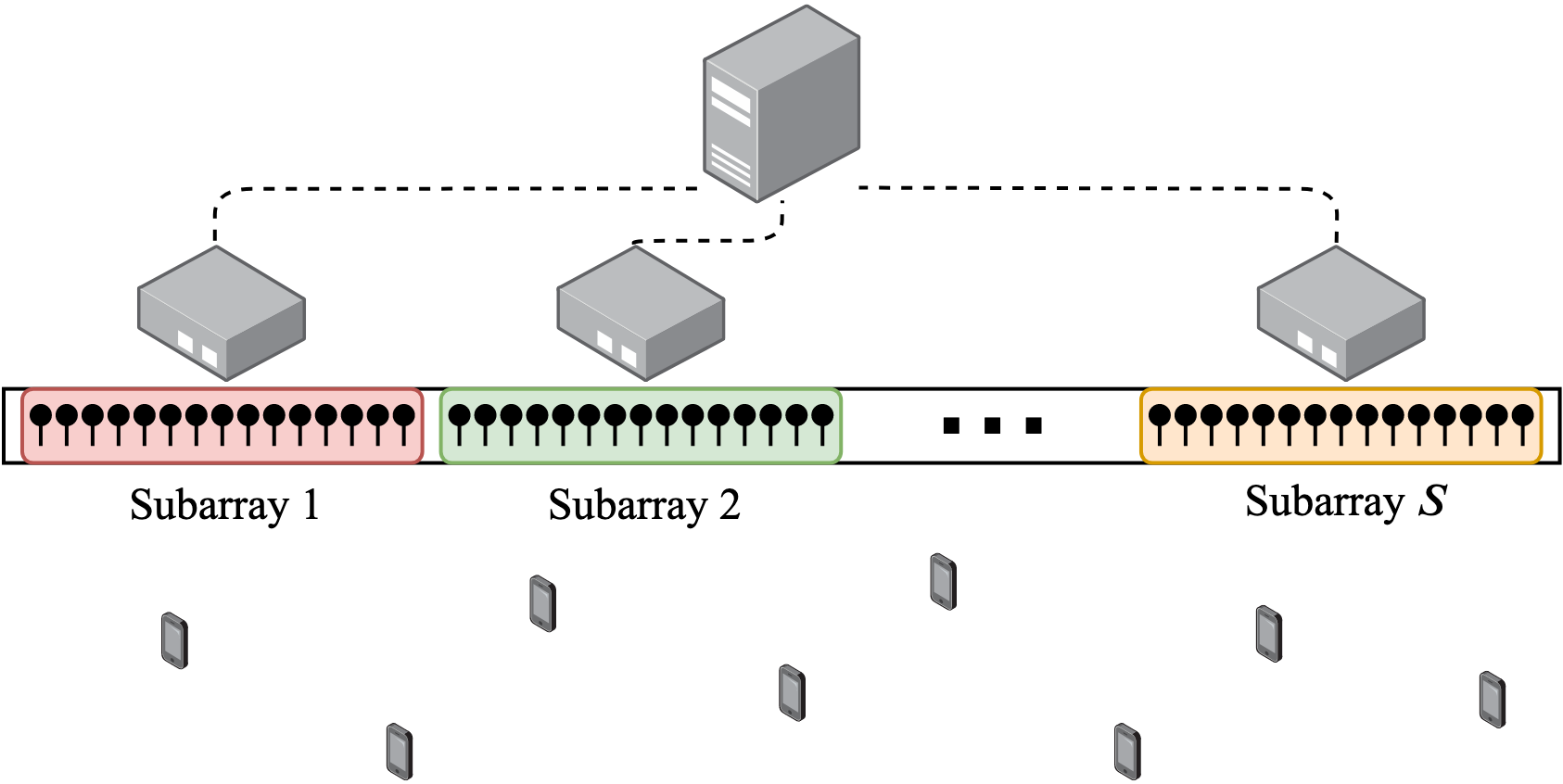}
	\caption{\small{XL-MIMO BS with fixed subarrays.}}
	\label{fig:II:01}
	\vspace{-6mm}
\end{figure}

In this section, we describe the uplink transmission phase of a XL-MIMO BS equipped with $M$ antennas that is serving $K$ single-antennas users. The users are using the same time-frequency resources and simultaneously transmitting data to the BS, where narrowband transmissions are considered. From now on, BS is supposed to know the channel state information (CSI) perfectly. {This communication setup is shown in Fig. \ref{fig:II:01}}.
Let $S$ be the number of fixed subarrays that splits an $M$ antenna array into disjoint groups of $M\sindex = M/S$ antennas, where $\sum^{S}_{s=1}M\sindex=M$ and each group has its own local processing unit for signal detection. A central unit is considered responsible for performing a data fusion operation {that combines the soft information} received by each subarray \cite{Amiri2019a}. Further, to ensure the benefits of M-MIMO, it is assumed that $M\sindex \geq K$. Thus, subarray $s$ receives the following baseband signal:
\begin{equation}
    \mathbf{y}\sindex = \sqrt{p}\mathbf{H}\sindex \mathbf{x} + \mathbf{n}\sindex,
    \label{eq:II:01}
\end{equation}
where $p$ is the uplink transmit power equal to all users, $\mathbf{H}\sindex\incplx{M\sindex\times{K}}=[\mathbf{h}\sindex_{1},\dots,\mathbf{h}\sindex_{K}]$ is the channel matrix of subarray $s$, $\mathbf{x}\incplx{K}$ contains the $K$ complex symbols messages with normalized power, and $\mathbf{n}\sindex \incplx{M\sindex} \sim \cplxgauss{\mathbf{0},\sigma^2\mathbf{I}}$ is a white Gaussian noise vector. Noise vectors are considered to be independent over the different subarrays.
The $M\sindex\times{1}$ channel vector with the channel coefficients of user $k$ to $M\sindex$ antennas of subarray $s$ is modeled as \cite{Amiri2019}
\begin{equation}
    \mathbf{h}\sindex_{k} = \sqrt{\mathbf{w}\sindex_{k}}\odot\bar{\mathbf{h}}\sindex_{k},
    \label{eq:II:02}
\end{equation}
where $\mathbf{w}\sindex_{k}$ embodies large-scale fading effects; 
Path-loss is modeled as$
    \mathbf{w}\sindex_{k} = \Omega \left(\mathbf{d}\sindex_{k}\right)^{-\nu}$,
where $\Omega$ is the path-loss attenuation coefficient, $\mathbf{d}\sindex_{k}\inreal{M\sindex}$ is a vector of the distances between user $k$ and each antenna of subarray $s$, and $\nu$ is the path-loss exponent. Channel effects resulting from small-scale fading are embraced by $\bar{\mathbf{h}}\sindex_{k} \sim \cplxgauss{\mathbf{0},\boldsymbol{\Theta}\sindex_{k}}$, where $\boldsymbol{\Theta}\sindex_{k}\inreal{M\sindex\times M\sindex}$ is the subarray channel covariance matrix that takes into account non-stationarity and spatial channel correlation effects.
The overall channel covariance matrix of the antenna array is then $\boldsymbol{\Theta}_{k}\inreal{M\times M} = \mathrm{blkdiag}(\boldsymbol{\Theta}^{(1)}_{k},\dots,\boldsymbol{\Theta}^{(S)}_{k})$ and
\begin{equation}
    \boldsymbol{\Theta}_{k} = \mathbf{D}^{\frac{1}{2}}_{k}\mathbf{R}_{k}\mathbf{D}^{\frac{1}{2}}_{k}, 
    \label{eq:II:04}
\end{equation}
where $\mathbf{R}_{k}\inreal{M\times M}$ is a symmetric positive semi-definite matrix that captures spatial channel correlation effects and $\mathbf{D}_{k}\in\{0,1\}^{M\times M}$ is a diagonal, indicator matrix that embraces non-stationary modeled through the VR concept.

\subsection{Visibility Regions (VRs)}\label{sec:sub:vrs}
The VRs describe the portion of the array being "viewed" by each user, i.e., where most portion of users' energy is concentrated. In particular, we adopt the model described in \cite{Amiri2019}, wherein each user has a VR identified by two main properties: its center and its length. Thus, VR centers are modeled as $c_{k}\sim\uniform{0,L}$, where $L$ is the XL-MIMO antenna array physical length, whereas VR lengths $l_{k}\sim{\mathcal{LN}}({\mu_{l},\sigma_{l}})$. 

Let denote the number of active antennas that are serving user $k$ as $D_{k}$, which is defined as the sum of antennas within the physical region delimited by $[c_{k}-l_{k},c_{k}+l_{k}]$. Hence, the diagonal matrix $\mathbf{D}_{k}$, introduced in \eqref{eq:II:04}, has $D_{k}$ non-zero diagonal elements. In the sequel, two different power normalizing schemes for the non-stationary channels are revisited \cite{Ali2019}.

\noindent
\textbf{Normalization 1.} Stationary and non-stationary channels have the same norm, i.e., $\trace{\nscovmatrix} = \trace{\mathbf{R}_{k}} = M \ \forall k$. This is achieved by  $\mathbf{D}_{k}=\diag{[\zeros{},(M/D_{k})^{1/2}\ones{D_{k}},\zeros{}]\transp}$.

\noindent
\textbf{Normalization 2.} Non-stationary channels have norm (in general) less than or equal to stationary ones. In this case, $\trace{\nscovmatrix} = D_{k} \ \forall k$ and $\mathbf{D}_{k}=\diag{[\zeros{},\ones{D_{k}},\zeros{}]\transp}$.

\subsection{Signal-to-Interference-Plus-Noise Ratio (SINR)}
Considering that the data symbols of each user are i.i.d. and Gaussian distributed, the instantaneous uplink SINR $\sinrks$ of user $k$ regarding subarray $s$ can be defined as:
\begin{equation}
    \sinrks = \frac{p\abs{(\vvectors{k})\herm\hvectors{k}}^{2}}{p\sum_{i=1,i\neq k}^{K} \abs{(\vvectors{k})\herm\hvectors{i}}^{2} + \sigma^2 \norm{\vvectors{k}}^2},
\end{equation}
where $\vvectors{k}\incplx{M\sindex}$ is the receive combining vector of subarray $s$. Recall that the objective of this work boils down to obtain an efficient way to compute $\vvectors{k}$ in terms of performance-complexity trade-off.

\section{Randomized Kaczmarz Signal Detection}
{The rKA is an iterative algorithm that solves systems of linear equations (SLEs) and has been recently applied to efficiently tackle the problem of relaxing linear signal processing schemes in the context of M-MIMO. This procedure was first presented in \cite{Boroujerdi2018} and deepened in \cite{Boroujerdi2019,Rodrigues2019}. The randomization in rKA is related to the order in which the SLE equations are being selected when solved. Modified and novel random selection methods that exploit non-stationary effects are discussed here.}

Each BS {with fixed} subarray {dimensions} is interested in detecting the users's transmitted symbols. In the context of M-MIMO, ZF and RZF are two widely used schemes that, for the sake of argument, can be applied over each fixed subarray, {yielding the following symbol estimates when $\xi=0$ (ZF) or $\xi\neq{0}$ (RZF):}
\begin{equation}
	\hat{\mathbf{x}}\sindex = \!(\mathbf{V}\sindex)\herm \mathbf{y}\sindex \!=\! [(\mathbf{H}\sindex)\herm\mathbf{H}\sindex\! + \xi \mathbf{I}_{K}]^{-1}\!(\mathbf{H}\sindex)\herm\mathbf{y}\sindex,
	\label{eq:III:B:01}
\end{equation}
where $\mathbf{V}\sindex\incplx{M\sindex\times{K}} = [\vvectors{1},\dots,\vvectors{K}]$ is the receive combining matrix associated with subarray $s$ and $\xi=\frac{1}{\mathrm{SNR}} = \frac{\sigma^2}{p}$.

The problem with adopting the procedure described in \eqref{eq:III:B:01} when considering extremely large arrays is the increased computational cost of the matrix inversion in crowded scenarios and its inherent scalability with the growing number of antennas and subarrays. To circumvent this high computational complexity and alleviate/decrease the hardware cost of each subarray's processing unit, our proposal is to obtain the symbol estimates at each subarray by still relying on the ZF and RZF methodologies, but instead of using the classical computation form in \eqref{eq:III:B:01}, we apply the rKA to obtain them. The main idea behind this is to realize that \eqref{eq:III:B:01} can be posed as the following optimization problem \cite{Boroujerdi2018}:
\begin{equation}
	\argmin_{\boldsymbol{\varrho}\sindex\incplx{K}} \lVert\mathbf{H}\sindex\boldsymbol{\varrho}\sindex-\mathbf{y}\sindex\rVert^{2}_{2} + \xi \lVert{\boldsymbol{\varrho}\sindex}\rVert^{2}_{2},
	\label{eq:III:B:02}
\end{equation}
which can be compactly written as
\begin{equation}
	\argmin_{\boldsymbol{\varrho}\sindex\incplx{K}} \lVert\BBmatrix\boldsymbol{\varrho}\sindex-\mathbf{y}\sindex_{0}\rVert^{2}_{2},
	\label{eq:III:B:02}
\end{equation}
where $\boldsymbol{\varrho}\sindex$ represents the symbol estimate at subarray $s$, $\BBmatrix\incplx{(M\sindex+K)\times K}=[\mathbf{H}\sindex;\sqrt{\xi}\mathbf{I}_{K}]$, while $\mathbf{y}\sindex_{0}\incplx{M\sindex+K} = [\mathbf{y}\sindex;\mathbf{0}]$. The symbol estimate vector in \eqref{eq:III:B:01} becomes
\begin{equation}
    \hat{\mathbf{x}}\sindex=[(\BBmatrix)\herm\BBmatrix]^{-1}(\BBmatrix)\herm\mathbf{y}\sindex_{0}.
    \label{eq:III:B:03}
\end{equation}

\subsection{Signal Estimates for Each Subarray via rKA}
To derive the rKA-based signal detection schemes at each subarray $s$, the key idea is to solve the optimization problem by finding the solution of the SLE $\BBmatrix\boldsymbol{\varrho}\sindex=\mathbf{y}\sindex_{0}$ via rKA, considering that each subarray is an independent MIMO system. However, due to the presence of arbitrary noise in the receive signal, it is possible to observe that this SLE is inconsistent, i.e., if the rKA is applied to solve this system, a high level of residual error would be obtained. To solve this problem, the authors of \cite{Boroujerdi2018} proposed a suitable transformation over the above SLE to remove the inconsistency, by solving the SLE in two steps: 
{\underline{Step.1}. Estimation} for ${\mathbf{y}}\sindex_{0}$ as
\begin{equation}
    \yHats=\BBmatrix\hat{\mathbf{x}}\sindex\overset{(a)}{=}\BBmatrix([\BBmatrix]\herm\BBmatrix)^{-1}(\mathbf{H}\sindex)\herm\mathbf{y}\sindex,
    \label{eq:III:B:04}
\end{equation}
where in ($a$) we used \eqref{eq:III:B:03}. Note that $\yHats$ lies in the subspace spanned by the columns of $\BBmatrix$. Thus, the following SLE can be obtained:
\begin{align}
      (\BBmatrix)\herm\yHats&=([\BBmatrix]\herm\BBmatrix)([\BBmatrix]\herm\BBmatrix)^{-1}(\mathbf{H}\sindex)\herm\mathbf{y}\sindex\\
      &= (\mathbf{H}\sindex)\herm\mathbf{y}\sindex, \nonumber \label{eq:III:B:05}
\end{align}
which can be written as:
\begin{equation}
    (\BBmatrix)\herm\mathbf{w}\sindex = \mathbf{b}\sindex
    \label{eq:III:B:06}
\end{equation}
where $\mathbf{w}\sindex\incplx{M\sindex+K}$ plays the role of $\yHats$ as an unknown vector, while $\mathbf{b}\sindex = (\mathbf{H}\sindex)\herm\mathbf{y}\sindex$. This SLE outputs $\yHats$ and represents the first step to obtain the signal estimates.

\noindent{\underline{Step.2}.} Without loss of generality, lets assume that $\yHats$ can be recovered through the solution of \eqref{eq:III:B:06} via rKA. With $\yHats$, the SLE in \eqref{eq:III:B:04} can be solved to obtain the estimates of the symbols transmitted by the users. This second SLE does not need to be solved directly, since, in the recover of $\yHats$, we can already obtain $\hat{\mathbf{x}}\sindex$ via the solution of $(\BBmatrix)\herm\mathbf{w}\sindex = \mathbf{b}\sindex$ by considering the $K$ last components of $\mathbf{w}$ divided by $\sqrt{\xi}$, where $\mathbf{b}\sindex = (\mathbf{H}\sindex)\herm\mathbf{y}\sindex$.

\subsection{Receive Combining Matrix for Each Subarray via rKA}
For scenarios where the channel coherence block is large, it turns out that the procedure described above is not computationally efficient, since we have to compute it to get estimates of $\hat{\mathbf{x}}\sindex$ at each complex-valued sample of the coherence block. A better way would be to have a method that computes $\mathbf{V}\sindex$ only once, and then use this information to compute all the signal estimates concerning a given coherence block\footnote{{This procedure, however, would not be adequate in cases where channel responses fluctuate rapidly.}}. The key to finding a way to get an estimate of the receive combining matrix $\hat{\mathbf{V}}\sindex$ is to note that is a scaled version of the $K$ receive combining vectors can be acquired when we have $K$ different SLEs of the form $(\BBmatrix)\herm\mathbf{w}\sindex_{i} = \mathbf{e}_{i}$, where $\mathbf{e}_{i}$ is the $i$th canonical basis, i.e., a vector comprised of zeros with a single value one in the $i$th position, for $i = 1,2,\dots,K$. It can be argued that this SLE results in a scaled estimate of the receive combining vector $\vvectors{i}$ of user $i$ (see further details in Section V of \cite{Boroujerdi2018}). As a result, if this SLE is solved for each user $i$, we can obtain an estimate of $\hat{\mathbf{V}}\sindex$, which can be used to get the symbol estimates $\hat{\mathbf{x}}\sindex$. These observations yield in the procedure summarized in Algorithm \ref{algo:01}. Note that the $K$ rKAs carried out by a subarray $s$ can be executed in parallel {in a commodity hardware}, i.e., they are independent, their randomness may or may not be shared\footnote{{The version of Algorithm \ref{algo:01} comes from \cite{Rodrigues2019}, which considers a self-initialization procedure to ensure and accelerate convergence for all users, i.e. both center- and edge-located users (see Step 10 of Algorithm \ref{algo:01}).}}, {and the processing can be distributed over cheap, not-so-powerful computing units.} 

\subsection{Algorithm Features and Data Fusion}
The main differences of Algorithm \ref{algo:01} for XL-MIMO in comparison to its analogous counterpart for M-MIMO are: (i) the algorithm does not need to run over users that do not have sufficient (or any) power present at subarray $s$, see step 5; this comes from the non-stationary nature of extremely large arrays which implies that users are only being served by a limited number of subarrays, and (ii) each subarray's distributed unit needs to execute the algorithm possibly with a different number of iterations $T\sindex$ for a central unit to get all symbol estimates {$\hat{\mathbf{x}}\sindex = (\hat{\mathbf{V}}\sindex)\herm\mathbf{y}\sindex$} for $s = 1,\dots,S$; then, the central unit applies a final data fusion step over these estimates to obtain a coherent detection of the symbols sent by all users across the different subarrays. {In Section \ref{sec:numResults}, we use the distributed linear data fusion (DLDF) receiver described in \cite{Amiri2019a}, which attempts to minimize the mean-squared error of users' signal estimates at each subarray.}

\subsection{Different Update Schedule Schemes for XL-MIMO}\label{sec:sub:diffways}

{In the context of rKA, the manner and order in which selection of the random rows occurs is often called as the \textit{update schedule}. The convergence speed of the rKA is closely tied to the updating schedule strategy, and this has motivated the study of randomized variants in new application scenarios, such as \cite{Boroujerdi2019}, \cite{Strohmer2006}. This basically translates into the choice of the probability vector $\mathbf{p}\sindex = [P\sindex_{1},\dots,P\sindex_{K}]\transp$ in step 12 of Algorithm \ref{algo:01}. Below, it is introduced some possible but effective ways to select the rows $r(t)$ in the context of XL-MIMO by trying to exploit the non-stationary properties. In particular, we present a novel approach, as well as alter different known ones in order to exploit non-stationary effects. One can note that all three strategies described in the sequel can be thought as different {\it power allocation} methods.}

\begin{algorithm}[!htbp]
	\centering
	\caption{\small {Receive Combining Matrix Estimation for Each Subarray using rKA}.}
	\label{algo:01}
	\small
	\begin{algorithmic}[1]
		\State \textbf{Input:} Number of subarray antennas $M\sindex$, number of users $K$, inverse of the SNR $\xi\geq0$ (RZF regularization factor), subarray channel matrix ${\mathbf{H}\sindex}\incplx{M\sindex\times{K}}$, and number of iterations $T\sindex$.
		\State \textbf{Initialization:} Specify $\mathbf{W}\sindex\incplx{K\times{K}} = \zeros{}$.
		\State \textbf{Procedure:}
		\For{$k \leftarrow 1$ {\bf to} $K$} 
		    \If {\textit{power of user $k$ is not zero}}
		    \State Define state vectors $\mathbf{u}^{t}\incplx{M\sindex}$ and $\mathbf{z}^{t}\incplx{K}$ with $\mathbf{u}^{0}=\zeros{}$ and $\mathbf{z}^{0}=\zeros{}$.
		    \State Define user canonical basis $\mathbf{e}_{k}\inreal{K}$, where $[\mathbf{e}_{k}]_{k}=1$ and $[\mathbf{e}_{k}]_{j}=0$, $\forall{j}\neq{k}$.
		    \For{$t \leftarrow 0$ {\bf to} {$T\sindex-1$}}
		        \If {$t = 0$}
		            \State Pick row $k$ of ${(\mathbf{H}\sindex})\herm$ as a way to coherently initialize the algorithm and make it fair. This is referred to as \textit{self-initialization} \cite{Rodrigues2019}.
		        \Else
		            \State Pick a row $r(t)$ of ${(\mathbf{H}\sindex})\herm$ with $r(t) \in \{1,2,\dots,K\}$ drawn based on $\mathbf{p}\sindex$ {(see Section \ref{sec:sub:diffways}).}
		        \EndIf
	        	\State Compute the residual: $$\eta^{t}:=\frac{[\mathbf{e}_{k}]_{r(t)}-\langle{\mathbf{h}}\sindex_{r(t)},\mathbf{u}^{t}\rangle - \xi z^{t}_{r(t)}}{\lVert{\mathbf{h}}\sindex_{r(t)}\rVert^{2}_{2} + \xi}.$$
	        	\State Update $\mathbf{u}^{t+1} = \mathbf{u}^{t} + \eta^{t}{\mathbf{h}}\sindex_{r(t)}$.
		        \State Update $z^{t+1}_{r(t)} = z^{t}_{r(t)} + \eta^{t}$.
		        \State Repeat $z^{t+1}_{j} = z^{t}_{j}, \ \forall j \neq r(t)$.
		    \EndFor 
		    \State Update $\left[\mathbf{W}\sindex\right]_{:,k}$ = $\mathbf{z}^{T\sindex-1}$.
		    \EndIf
		\EndFor
		\State \textbf{Output:} $\mathbf{W}\sindex$, $\hat{\mathbf{V}}\sindex={\mathbf{H}\sindex}\mathbf{W}\sindex$.
	\end{algorithmic}
\end{algorithm}

\subsubsection{Power-based update schedule (pwr.)} The traditional rKA sample probability in the context of Algorithm \ref{algo:01} {is \cite{Strohmer2006}}
\begin{equation}
    P\sindex_{r(t)}= \frac{\lVert{\mathbf{h}}\sindex_{r(t)}\rVert^{2}_{2} + \xi}{\lVert{\mathbf{H}\sindex}\rVert^{2}_{\mathrm{F}} + K\xi}.
    \label{eq:III:C:01}
\end{equation}
This probability can be interpreted as the relative ratio of the power of user $r(t)\in\{1,\dots,K\}$ to the power of all users in the system. Therefore, users with better channel conditions or/and \textit{now} with more active antennas $D_{k}$ at a specific subarray $s$ are more often chosen. Moreover, to compute this sample probability, we need to obtain the $K$ sample probabilities of each user in which each takes $2M\sindex$ complex multiplications \cite[Appx. B]{Bjornson2017}. In fact, due to non-stationary, not all users will be served by subarray $s$, and therefore only $\meanK$ samples probabilities need to be computed, where $\meanK$ is the average number of users served by each subarray.

\subsubsection{Uniform update schedule (unif.)}
A second {strategy} for the sample probability was suggested by the authors in \cite{Boroujerdi2019}. The authors of \cite{Boroujerdi2019} proved that, if the selection of the rows is defined to be uniform with respect to the users i.e., $P\sindex_{r(t)} = 1/K\sindex$, the rKA also achieves an expected rate of convergence, where $K\sindex$ denotes the number of active users at subarray $s$. This method can be considered to bring fairness to the update schedule, in the sense that no user-specific equations are preferable. Different from the previous case, we assume that no extra computational complexity is required to compute $\mathbf{p}\sindex$. 

\subsubsection{Active-antennas-based update schedule (a.a.)} Aiming the exploitation of non-stationary channels, {herein, we propose} an update schedule scheme which is similar to the uniform one, but now the samples probabilities are based on the number of active antennas $D\sindex_{k}$ of user $k$ at subarray $s$. We define the sample probability as
\vspace{-4mm}

\begin{equation}
    P\sindex_{r(t)} = \frac{D\sindex_{r(t)}}{\sum_{i=1}^{K\sindex}D\sindex_{i}}.
\end{equation}
This approach gives more attention to users that have a large number of active antennas at each subarray. Again, no additional computational complexity is considered.

\section{Computational Complexity Analysis}
In this section, we characterize the computational complexity of Algorithm \ref{algo:01}. To do so, we consider the framework for complexity analysis presented in \cite[Appx. B]{Bjornson2017}, where only complex multiplications/divisions are taken into account.

Table \ref{tab:ULcomplexity} summarizes the computational complexity expressions of the traditional ZF and RZF schemes \cite{Bjornson2017}, as benchmarks, and of Algorithm \ref{algo:01} when considering the three different update schedule schemes discussed in Section \ref{sec:sub:diffways}. 

{Some observations are now in order:
\begin{enumerate}
    \item We consider that the vector norms $\lVert\mathbf{h}\sindex_{r(t)}\rVert_{2}$ are computed once and then they are stored at each subarray's processing unit. The only other operation that contributes to the computational complexity is $\langle{\mathbf{h}}\sindex_{r(t)},\mathbf{u}^{t}\rangle$ at each iteration $t$.
    \item The reception columns refers to the computation of $\hat{\mathbf{x}}\sindex = (\mathbf{V}\sindex)\herm\mathbf{y}\sindex$ at each subarray. From the point of view of low-complexity, we can maintain the output of Algorithm \ref{algo:01} in the factorized form $\mathbf{W}\sindex$ and perform
    $\hat{\mathbf{x}}\sindex = ([\mathbf{W}\sindex]\herm([\mathbf{H}\sindex]\herm\mathbf{y}\sindex))$ to recover the symbol estimates at each complex-valued sample of the coherence block. Given that the number of complex-valued samples reserved to the uplink phase is $\tau_{\mathrm{ul}}$, the above operation leads to $\tau_{\mathrm{ul}}M\sindex \meanK$ complex multiplications at each subarray distributed unit.
    \item We assume that both the canonical in \eqref{eq:III:B:01} and rKA forms of computing the ZF and RZF receive combining matrices are taking advantage of the non-stationary premise that not all users are served by all subarrays. 
    \item The overall computational complexity is given by the computation of all $\hat{\mathbf{x}}\sindex$'s, where it is important to note that the number of iterations may vary for each subarray. 
\end{enumerate}
}
\begin{table}[!htbp]
	\centering
	\caption{\small{Overall computational complexity per coherence block for the XL-MIMO receive combining schemes based on complex operations}}
	\label{tab:ULcomplexity}
	\scalebox{0.9}{
	\begin{tabular}{|p{1cm}|p{3cm}|p{1cm}|p{2cm}|}
		\hline
		\multirow{2}{*}{{\textbf{Scheme}}} & \multicolumn{2}{|c|}{{\textbf{Receive combining matrix}}} & {\textbf{Reception}} \\ \cline{2-4} 
		& {\textit{Multiplications}} & {\textit{Divisions}} & {\textit{Multiplications}} \\ \hline
		ZF  & $S[({3(\meanK)^{2}M\sindex})/{2}+({\meanK M\sindex})/{2}+({(\meanK)^3-\meanK})/{3}]$  & $S\meanK$ & $\tau_{\mathrm{ul}}SM\sindex\meanK$ \\ \hline
		RZF & $S[({3(\meanK)^{2}M^{(s)}})/{2}+({3\meanK M^{(s)}})/{2}+({(\meanK)^3-\meanK})/{3}]$ & $S\meanK$ & $\tau_{\mathrm{ul}}SM^{(s)}\meanK$ \\ \hline
		{Alg. 1 (pwr.)} & $S[M\sindex T^{(s)}+2M^{(s)}\meanK]$ &  & $\tau_{\mathrm{ul}}SM^{(s)}\meanK$ \\ \hline
		{Alg. 1 (unif./a.a)} & $S[M\sindex T^{(s)}+M^{(s)}]$ &  & $\tau_{\mathrm{ul}}SM^{(s)}\meanK$ \\ \hline
	\end{tabular}
	}
\end{table}
\vspace{-2mm}

\subsection{Deriving Upper Bounds for the Number of Iterations}
From Table \ref{tab:ULcomplexity}, one can note that the computational advantage of the rKA in Algorithm \ref{algo:01} basically depends on the amount of iterations $T\sindex$ required for the algorithm to achieve a given convergence notion (an iterative stopping criterion). We now derive upper bounds for $T\sindex$ in the sense that, if the average number of iterations required to reach a given convergence notion exceeds these bounds, Algorithm \ref{algo:01} would perform worse than the canonical form of computing the ZF and RZF schemes, given in \eqref{eq:III:B:01}. In fact, without loss of generality, we focus only on the RZF scheme from now on\footnote{{As discussed in \cite{Boroujerdi2018}, \cite{Rodrigues2019}, most promising results are obtained for the RZF scheme due to the fact that the regularization factor $\xi$ assists in the convergence of the algorithm.}}. Comparing the rows in Table \ref{tab:ULcomplexity} and isolating the number of iterations, we have 
\begin{align}
\hspace{4mm}
\small T^{(s),\rm up}_{\rm pwr.} &= \frac{1}{3} \frac{(\meanK)^3}{M^{(s)}} + \frac{2}{3} \frac{\meanK}{M^{(s)}} + \frac{3}{2} (\meanK)^2 - \frac{1}{2} \meanK \label{eq:IV:B:01}\\ 
\hspace{-2mm}T^{(s),\rm up}_{\rm unif.,a.a.} &= \frac{1}{3} \frac{(\meanK)^3}{M^{(s)}} + \frac{2}{3} \frac{\meanK}{M^{(s)}} + \frac{3}{2} (\meanK)^2 \nonumber \\ &+ \frac{3}{2} \meanK - 1 \label{eq:IV:B:02}.
\end{align}
These upper bounds are used in the convergence analysis carried out in Section \ref{sec:numResults:convergence}.

\section{Numerical Results {and Discussion}}\label{sec:numResults}
To verify the efficiency of Algorithm \ref{algo:01} in achieving a good performance-complexity trade-off solution for XL-MIMO signal detection, we now collect some quantitative results. The simulation parameters are disposed in Table \ref{tab:simulation-parameters}. The users are uniformly distributed inside a square-cell area with a minimum distance of $30$ m to the BS. The extremely large array follows a uniform linear array (ULA) arrangement with spacing between antennas of {$2\lambda$} m.

\begin{table}
    \centering
    \caption{\small{Simulation parameters.}}
    \label{tab:simulation-parameters}
    \scalebox{.9}{
    \begin{tabular}{cc|cc}
    \hline
    \textbf{Parameter} & \textbf{Value} & \textbf{Parameter} & \textbf{Value} \\ \hline
    Cell area & $0.1\times0.1$ km\textsuperscript{2} & Min. distance & $30$ m \\
    $M$ & $100$ & Array type & ULA \\
    $S$ & $4$ & Carrier frequency & $2.6$ GHz \\
    $M^{(s)}$ & $25$ & Antenna spacing & $2\lambda$ m \\
    $K$ & $25$ & $L$ & $23.0610$ m \\
    $p$ & $0$ dBm & Channel model & $\mathbf{R}_{k} = \eye{M}$ \\
    $\sigma^2$ & $[-55,-40]$ dBm & $c_{k}$ & $\mathcal{U}(0,L)$\\
    $\Omega$ & $4$ & $l_{k}$ & $\mathcal{LN}(0.1L,0.1)$ \\
    $\nu$ & $3$ &  & \\ \hline
    \end{tabular}%
    }
\end{table}

\subsection{Convergence Analysis}\label{sec:numResults:convergence}
{Here, we characterize SNR regions in which the proposed algorithm brings relevant computational gains.} To ease the exposition, we define the following quantity called as the \textit{computational relaxation degree} $\textsc{crd}\sindex_{i}$:
\begin{equation}
    \textsc{crd}\sindex_{i} =  \frac{T^{(s),\rm up}_{i} - \bar{T}\sindex}{T^{(s),\rm up}_{i}}, \text{ if } \bar{T}\sindex <  T^{(s),\rm up}_{i}
\end{equation}
and $0$ otherwise, where $\bar{T}\sindex$ is the average number of iterations per subarray needed to achieve a sense of appropriate convergence and $i$ indexes the different update schedules. This quantity measures the relative computational complexity gains obtained for each subarray via Algorithm \ref{algo:01} compared to the canonical way of computing the RZF scheme.

\begin{figure}[!htbp]
	\centering
	\subfloat[Average $\textsc{crd}$ $\times$ noise variance in dBm.]{%
		\includegraphics[width=.48\textwidth]{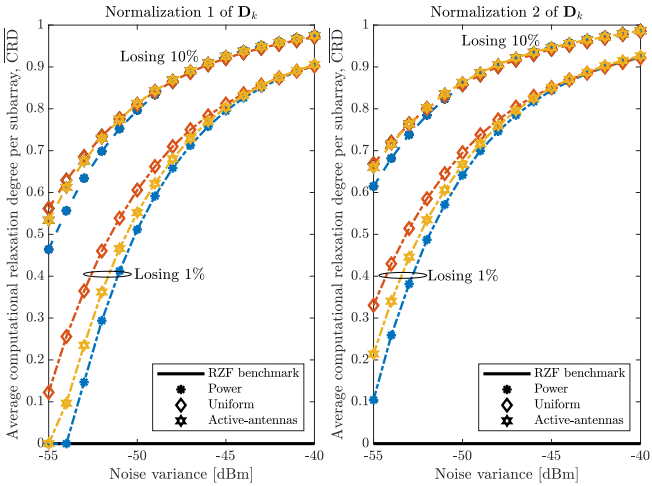}%
		\label{fig:V:A:01:a}
	}
	
	
	\subfloat[Average SER $\times$ noise variance in dBm.]{%
		\includegraphics[width=.48\textwidth]{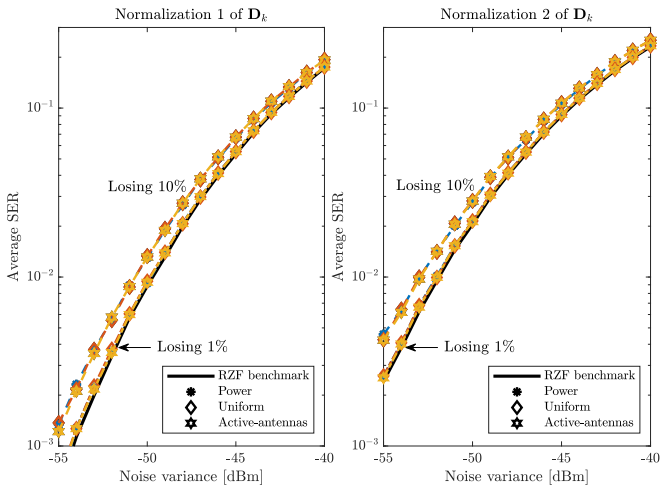}%
		\label{fig:V:A:01:b}
	}
	
	\caption{\small{Performance-complexity trade-off. $K/M = 0.25$ and $p=0$ dBm. Performance gaps of 10\% and 1\% regarding the canonical RZF scheme and two ways of normalizing $\mathbf{D}_{k}$.}}
	\label{fig:V:A:01}
\end{figure}

\begin{figure}[!htbp]
	\centering
	\includegraphics[width=.48\textwidth]{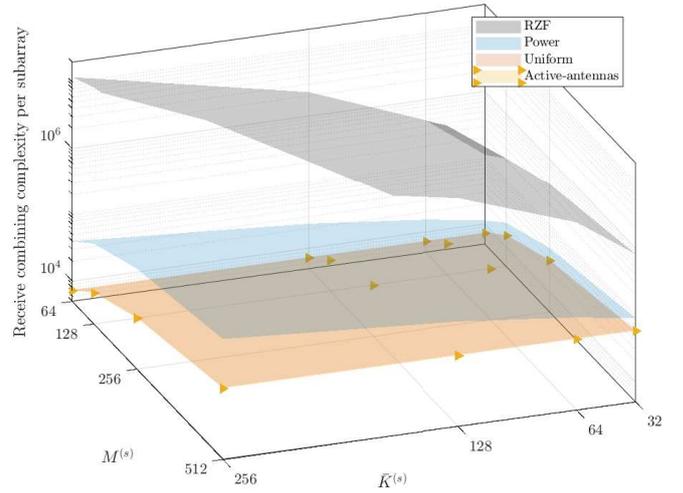}
	\caption{\small{Receive combining computational complexity as a function of $M\sindex$ and $\meanK$. $p = 0$ dBm, $\sigma^2 = 48$ dBm, normalization 2 is considered, and the number of iterations is fixed according to Fig. \ref{fig:V:A:01:a} based on losing 10\%.}}
	\label{fig:V:A:02}
\end{figure}

Fig. \ref{fig:V:A:01:a} shows the computational relaxation degree as a function of the noise variance in dBm. Note that both ways of normalizing $\mathbf{D}_{k}$ discussed in Section \ref{sec:sub:vrs} were considered. The average number of iterations were obtained by comparing the average SINR of Algorithm \ref{algo:01} with the average SINR benchmark given by the canonical computation of RZF at each subarray. Moreover, two stopping criteria were considered in relation to the performance measured via average SINR: (i) Algorithm \ref{algo:01} outputs an estimate of $\mathbf{V}\sindex$ that reduces $10\%$ of the canonical performance of the RZF scheme, and (ii) the same but considering a losing in performance of only $1\%$. 
{We now made some observations:
\begin{enumerate}
    \item \textit{Average system performance:} uniform update schedule outperforms all the other schemes. This is because, for users with good and bad channel conditions, the algorithm converges properly.
    \item Active-antennas-based update schedule performs marginally better than the typical power-based one and has a considerably easier implementation.
    \item Normalization 2 better accelerates the algorithm convergence because of the disparity among the power of the users which reduces the overall average signal-to-interference ratio (SIR).
\end{enumerate}
The most important conclusion is that we can roughly resemble the performance provided by RZF by greatly reducing the computational complexity. At low SNR, this is easily achieved due to low interference among users. 

{Fig. \ref{fig:V:A:02} illustrates the relaxation in computational complexity brought by the algorithm when considering different system sizes. Note that the RZF complexity has a rapid growth in comparison with the rKA-based schemes as $M\sindex$ and $\meanK$ increase. Uniform and active-antennas approaches are the most attractive ones.}
}

\subsection{Performance Comparison}
To give a notion of the performance gap  of the two different adopted stopping criteria, Fig. \ref{fig:V:A:01:b} shows the average symbol error rate (SER) as a function of the noise variance in dBm. The number of iterations used for each different noise variance point follows the results obtained in Fig. \ref{fig:V:A:01:a}. To fusion the signal estimates of each subarray, we used the DLDF receiver described in \cite[Algorithm 1]{Amiri2019a}. It is important to observe that, although the algorithm rate of convergence when using Normalization 2 of $\mathbf{D}_{k}$ is faster (see Fig. \ref{fig:V:A:01:a}), its performance is impaired by the different array gains for each user in comparison with the first normalization method.\\

\section{{Conclusions}}\label{sec:conc}
In this work we have proposed a rKA-based combiner specifically applicable to XL-MIMO systems aiming at reducing the computational burden of the signal detector  with improved performance-complexity trade-off. We have provided a computational complexity analysis via upper bounds derivation for the number of iterations to achieve convergence (with 10\% or 1\% losing). Besides, we have proposed a new {\it update scheduler} for the rKA, namely active-antenna-based update schedule, aiming at exploiting the intrinsic  non-stationary  properties in XL-MIMO channels. Future  research  will address optimizing the complexity of systems with different user requirements.


\bibliographystyle{IEEEtran}

\begin{thebibliography}{10}
\providecommand{\url}[1]{#1}
\csname url@samestyle\endcsname
\providecommand{\newblock}{\relax}
\providecommand{\bibinfo}[2]{#2}
\providecommand{\BIBentrySTDinterwordspacing}{\spaceskip=0pt\relax}
\providecommand{\BIBentryALTinterwordstretchfactor}{4}
\providecommand{\BIBentryALTinterwordspacing}{\spaceskip=\fontdimen2\font plus
\BIBentryALTinterwordstretchfactor\fontdimen3\font minus
  \fontdimen4\font\relax}
\providecommand{\BIBforeignlanguage}[2]{{%
\expandafter\ifx\csname l@#1\endcsname\relax
\typeout{** WARNING: IEEEtran.bst: No hyphenation pattern has been}%
\typeout{** loaded for the language `#1'. Using the pattern for}%
\typeout{** the default language instead.}%
\else
\language=\csname l@#1\endcsname
\fi
#2}}
\providecommand{\BIBdecl}{\relax}
\BIBdecl

\bibitem{Bjornson2019}
E.~Bj{\"{o}}rnson, L.~Sanguinetti, H.~Wymeersch, J.~Hoydis, and T.~L. Marzetta,
  ``{Massive MIMO is a reality—What is next?: Five promising research
  directions for antenna arrays},'' \emph{Digital Signal Processing: A Review
  Journal}, vol.~94, pp. 3--20, 2019.

\bibitem{DeCarvalho2019}
E.~{De Carvalho}, A.~Ali, A.~Amiri, M.~Angjelichinoski, and R.~W. Heath,
  ``{Non-Stationarities in Extra-Large Scale Massive MIMO},'' mar 2019.

\bibitem{Bjornson2017}
E.~Bj{\"{o}}rnson, J.~Hoydis, and L.~Sanguinetti, ``{Massive MIMO Networks:
  Spectral, Energy, and Hardware Efficiency},'' \emph{Foundations and
  Trends{\textregistered} in Signal Processing}, vol.~11, no. 3-4, pp.
  154--655, 2017.

\bibitem{mller2013linear}
A.~Müller, A.~Kammoun, E.~Björnson, and M.~Debbah, ``Linear precoding based
  on polynomial expansion: Reducing complexity in massive mimo,'' 2013.

\bibitem{Boroujerdi2018}
M.~N. Boroujerdi, S.~Haghighatshoar, and G.~Caire, ``{Low-Complexity
  Statistically Robust Precoder/Detector Computation for Massive MIMO
  Systems},'' \emph{IEEE Transactions on Wireless Communications}, vol.~17,
  no.~10, pp. 6516--6530, oct 2018.

\bibitem{Amiri2019}
A.~Amiri, C.~N. Manch'on, and E.~de~Carvalho, ``A message passing based
  receiver for extra-large scale mimo,'' \emph{arXiv preprint
  arXiv:1912.04131}, 2019.

\bibitem{Boroujerdi2019}
M.~N. Boroujerdi, A.~Abbasfar, and M.~Ghanbari, ``{Efficient beamforming scheme
  in distributed massive MIMO system},'' \emph{International Symposium on Turbo
  Codes and Iterative Information Processing, ISTC}, vol. 2018-Decem, pp. 1--5,
  2019.

\bibitem{Rodrigues2019}
V.~C. Rodrigues, J.~C. {Marinello Filho}, and T.~Abr{\~{a}}o, ``{Randomized
  Kaczmarz algorithm for massive MIMO systems with channel estimation and
  spatial correlation},'' \emph{International Journal of Communication
  Systems}, p. e4158, sep 2019.

\bibitem{Ali2019}
A.~Ali, E.~{De Carvalho}, and R.~W. Heath, ``{Linear Receivers in
  Non-Stationary Massive MIMO Channels with Visibility Regions},'' \emph{IEEE
  Wireless Communications Letters}, vol.~8, no.~3, pp. 885--888, 2019.

\bibitem{Amiri2019a}
A.~Amiri, M.~Angjelichinoski, E.~{De Carvalho}, and R.~W. Heath, ``{Extremely
  Large Aperture Massive MIMO: Low Complexity Receiver Architectures},''
  \emph{2018 IEEE Globecom Workshops, GC Wkshps 2018 - Proceedings}, 2019.

\bibitem{Strohmer2006}
T.~Strohmer and R.~Vershynin, ``{A Randomized Solver for Linear Systems with
  Exponential Convergence},'' 2006, pp. 499--507.

\end{thebibliography}


\end{document}